\documentclass[epj]{svjour}
\usepackage{epsfig,amsmath,amssymb,scalefnt}

\newcommand{\abbrev}{\scalefont{.9}}

\newcommand{\eqn}[1]{Eq.\,(\ref{#1})}
\newcommand{\fig}[1]{Fig.\,\ref{#1}}

\newcommand{\lo}{{\abbrev LO}}
\newcommand{\nlo}{{\abbrev NLO}}
\newcommand{\nnlo}{{\abbrev NNLO}}

\newcommand{\reference}[1]{Ref.\,\cite{#1}}

\newcommand{\lhc}{{\abbrev LHC}}
\newcommand{\susy}{{\abbrev SUSY}}
\newcommand{\mssm}{{\abbrev MSSM}}
\newcommand{\sm}{{\abbrev SM}}
\newcommand{\qcd}{{\abbrev QCD}}
\newcommand{\msusy}[1]{m_{\tilde{#1}}}
\newlength{\figsize}
\begin{document}
\title{Supersymmetric Higgs production at the Large Hadron Collider
}
\authorrunning{Robert Harlander}
\author{Robert Harlander\thanks{Supported by DFG, contract HA\,2990/2-1.}
\\[-10em]
\hspace*{29em}{October 2003 --- TTP03-33 --- hep-ph/0311005}\\[8em]
}
\institute{Institut f\"ur Theoretische Teilchenphysik, Universit\"at
  Karlsruhe, 76128 Karlsruhe, Germany}
\date{}
\abstract{
We review the status of theoretical predictions for the production of
neutral Higgs bosons at the \lhc{}. Special emphasis is put on the
role of bottom quarks in the gluon fusion process and in the
associated production of Higgs bosons with $b\bar b$ pairs.
 % end of PACS codes
} %end of abstract
\maketitle

\section{Introduction}
\label{intro}
There has been significant progress in controlling the theoretical
predictions for Standard Model Higgs production cross sections at the
\lhc{} over the past few years. The Higgs sector may play a crucial role
for the discrimination among the Standard Model and various extended theories
at the \lhc{}. In the following, we discuss the status of theoretical
predictions for supersymmetric Higgs production, focussing on the
{\abbrev CP}-even neutral Higgs bosons, generically denoted by $H$.

\section{Gluon fusion}
\begin{figure}
  \begin{center}
    \leavevmode
    \begin{tabular}{cc}
      \epsfxsize=9em
      \raisebox{0em}{\epsffile[110 520 350 690]{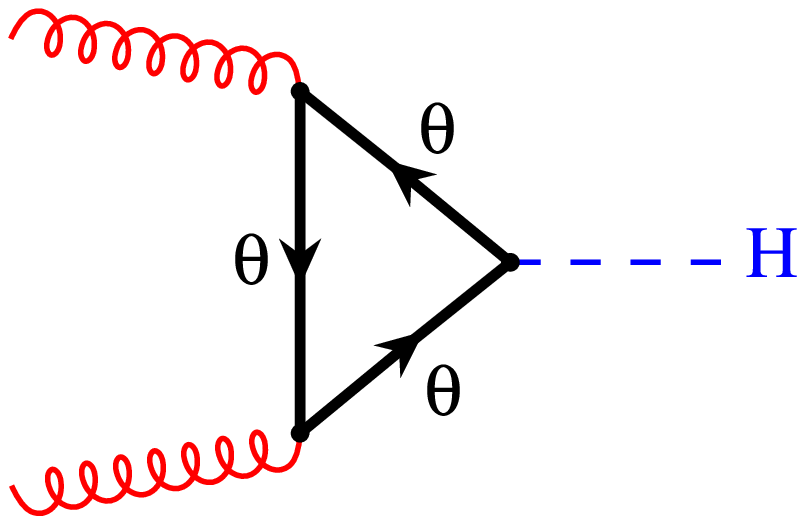}} &
      \epsfxsize=9em
      \raisebox{0em}{\epsffile[110 520 350 690]{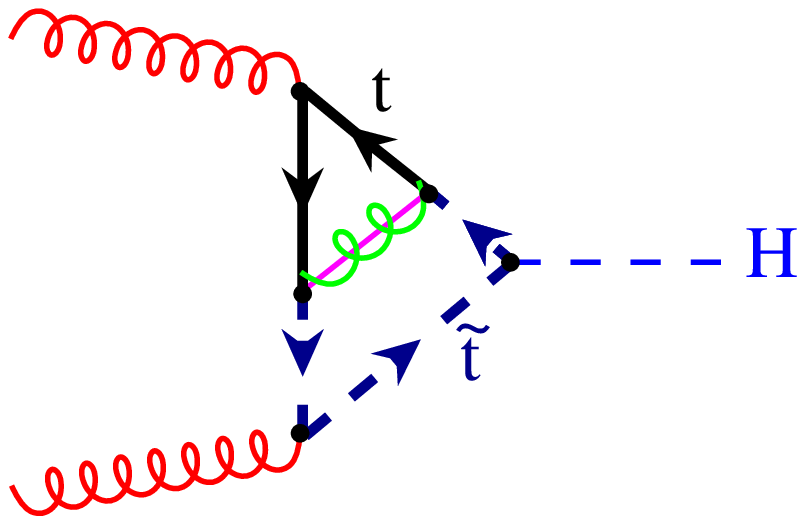}} \\
      $(a)$ & $(b)$
    \end{tabular}
      \caption[]{\label{fig::ggh}\sloppy
        Feynman diagrams contributing to gluon fusion:
      $(a)$~$\theta=t,b,\tilde t,\ldots$ --- $(b)$~two-loop contribution in
        the \mssm{}.}
  \end{center}
\end{figure}

For the \sm{} and most of its extensions, gluon fusion is the dominant
production mode for a neutral Higgs boson at hadron colliders.  A
generic \lo{} diagram is shown in \fig{fig::ggh}\,$(a)$.  The coupling
of the gluons to the Higgs boson is a pure quantum effect: In the \sm{},
it is mediated predominantly by top quarks, while the contribution of
other fermions $f$ is suppressed by $m_f/m_t$.

In the \mssm{}, the contribution from virtual bottom quarks can be
significantly enhanced through large values of $\tan\beta$. Furthermore,
top squarks can give a sizable contribution if their masses are of the
order of $m_t$.  We thus write the total cross section from gluon fusion
as
\begin{equation}
\begin{split}
\sigma = \sigma_t+\Delta\sigma_{b}+\Delta\sigma_{\tilde t}
+\Delta\sigma_{b\tilde t}\,,
\label{eq::sigdef}
\end{split}
\end{equation}
where $\sigma_t$ denotes the pure top quark contribution, and
$\Delta\sigma_{b}$ ($\Delta\sigma_{\tilde t}$) the additional effects
due to the presence of bottom quarks (top squarks). The term
$\Delta\sigma_{b\tilde t}$ which arises from the interference of bottom
quarks and top squarks, will not be considered any further in this
paper.  Note that we also neglect effects from bottom squarks which are
suppressed by $(m_b/m_{\tilde b})^2$, modulo a possible enhancement due
to large values of $\tan\beta$.  Furthermore, we define
\begin{equation}
\begin{split}
\sigma_{ti} \equiv \sigma_t + \Delta\sigma_{i}\,,\qquad 
i\in \{b,\tilde t\}\,.
\end{split}
\end{equation}

\begin{figure}
  \begin{center}
    \leavevmode
    \begin{tabular}{c}
      \epsfxsize=\figsize
      \epsffile[110 265 465 560]{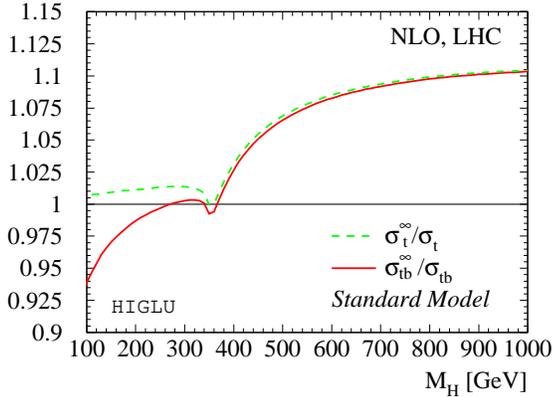}
    \end{tabular}
      \caption[]{\label{fig::efft_lhc}\sloppy The total cross section at
	\nlo{} as evaluated in the effective theory (\eqn{eq::efft}),
	compared to the exact \nlo{}
	result~\cite{Spira:1995rr,higlu}. Dashed line: only top quarks
	--- solid line: including bottom quarks ($m_t^{\rm OS} =
	175$\,GeV, $m_b^{\rm OS} = 5$\,GeV).}
  \end{center}
\end{figure}

\paragraph{Top quark loops, Effective theory approach.}
Apart from an overall factor $f(\beta)$, the pure top quark
contributions $\sigma_t$ to the Higgs production cross section in the
\mssm{} and the \sm{} are identical to all orders in \qcd{}. Higher
orders are described extremely well by the ``heavy-top
limit''~\cite{nlo} which, for this particular case, is defined as
follows:
\begin{equation}
\begin{split}
  \sigma_\theta^{\infty} &= \kappa_\theta\cdot
  \sigma_\theta^{(0)}\,,\qquad \kappa_\theta =
  \frac{\sigma_\theta(m_\theta\to\infty)}{\sigma_\theta^{(0)}(m_\theta\to
  \infty)}\,,
  \label{eq::efft}
\end{split}
\end{equation}
where $\theta=t$. Here, $\sigma_\theta^{(0)}$ denotes that
part of $\sigma_\theta$ where all higher order corrections to the {\sl
partonic} process are dropped (they are kept in the parton densities and
the running of $\alpha_s$). \fig{fig::efft_lhc} shows that the
difference between $\sigma^{\infty}_t$ and the exact result $\sigma_t$
is less than 2\% for $M_H<2m_t$ at next-to-leading order. This should be
compared to the \nlo{} uncertainty of $\pm 15\%$~\cite{nlo} as estimated
from varying the renormalization and factorization scales $\mu_R$ and
$\mu_F$.  It was therefore well justified to apply the heavy-top limit
at \nnlo{}~\cite{Harlander:2000mg,gghnnlo}\footnote{For resummation
effects, see \reference{gghresum}.} in order to obtain a theoretical
prediction for the total cross section that is competitive with the
expected experimental uncertainty.\\[-2em]

\begin{figure}
  \begin{center}
    \leavevmode
    \begin{tabular}{c}
      \epsfxsize=\figsize
      \epsffile[110 265 465 560]{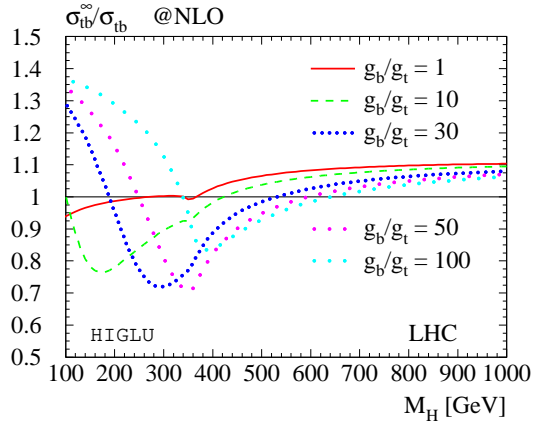}
    \end{tabular}
      \caption[]{\label{fig::btanb_lhc}\sloppy Relevance of the exact
	bottom quark contribution for various values of the bottom
	Yukawa coupling~\cite{higlu}.  $g_b/g_t=1$ corresponds to the
	Standard Model (see also~\cite{bill}).  }
  \end{center}
\end{figure}

\paragraph{Bottom quark loops.}
As was mentioned above, the gluon-Higgs coupling gets significant
contributions from bottom quark loops for large values of $\tan\beta$.
Since the ``heavy-top limit'' works at the 10\% level even for very
large Higgs boson masses (see \fig{fig::efft_lhc}), it is tempting to
apply a formal ``heavy-bottom approach'', defined by \eqn{eq::efft} with
$\theta=tb$ and $m_{tb}\equiv \{m_t,m_b\}$.  At \nlo{}, it is
$\kappa_{tb}=\kappa_{t}$.  \fig{fig::btanb_lhc} shows the deviation of
$\sigma_{tb}^\infty$ from the exact result at
\nlo{}~\cite{Spira:1995rr,higlu} for various values of the ratio
$g_b/g_t$, where $g_{b,t}$ are the Yukawa couplings of the bottom and
top quark {\sl relative to their \sm{} values}.  Note that the solid/red
curves (Standard Model) of Figs.\,\ref{fig::btanb_lhc}
and~\ref{fig::efft_lhc} are identical.

The curves in \fig{fig::btanb_lhc} show that the effect of the exact
\nlo{} bottom contribution stays below 40\% even for very large bottom
Yukawa couplings. For large Higgs boson masses, the curves approach the
Standard Model value (solid/red curve).\\[-2em]

\paragraph{SUSY loops.}
\begin{figure}
  \begin{center}
    \leavevmode
    \begin{tabular}{c}
      \epsfxsize=\figsize
      \epsffile{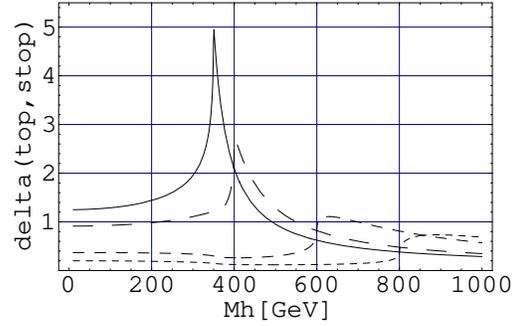}
    \end{tabular}
      \caption[]{\label{fig::top-stop}\sloppy
	Relative size of the top quark/squark contributions:
	{\tt delta(top,stop)=}$\Delta\sigma_{\tilde t}/\sigma_t$,
	see\,\eqn{eq::sigdef}. Furthermore, $m_t=175$\,GeV, and
	$m_{\tilde tR}=m_{\tilde tL}\equiv m_{\tilde t}$.
	Solid line: $m_{\tilde t}=175$\,GeV ---
	long/middle/short dashes: $m_{\tilde t} = 200/300/400$\,GeV.
      }
  \end{center}
\end{figure}
The contribution of squarks to the total Higgs production cross section
goes like $(m_q/\msusy{q})^2$. Thus, as shown in \fig{fig::top-stop},
only top squarks with $\msusy{t}\lesssim 400$\,GeV give a sizable
effect.

The \susy{} relation between the top and stop Yukawa coupling requires
to include also gluino effects at higher orders in $\alpha_s$
to arrive at finite results. A sample diagram with top quark,
top squark, and gluino is displayed in \fig{fig::ggh}\,$(b)$.

The \nlo{} corrections (evaluated through \eqn{eq::efft} with
$\theta=t\tilde t$ and $m_{t\tilde t} \equiv \{m_t,m_{\tilde t},
m_{\tilde g}\}$) were found to be very similar to the Standard Model
case~\cite{Harlander:2003bb} (see also \reference{Dawson:1996xz}, so
that the tree-level ratios shown in \fig{fig::top-stop} hardly change at
\nlo{}. In this first study, squark mixing effects had been neglected,
but more detailed investigations are under way.

\begin{figure}
  \begin{center}
    \leavevmode
    \begin{tabular}{c}
      \epsfxsize=\figsize
      \epsffile[110 265 465 560]{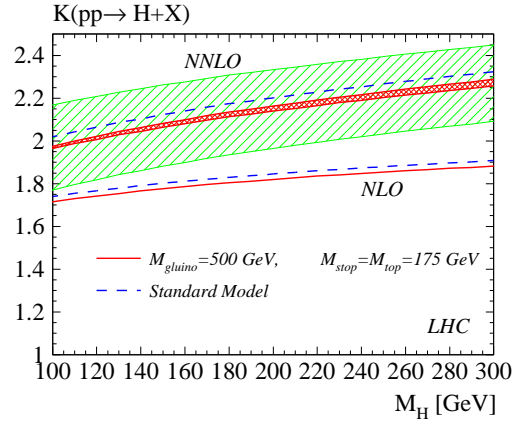}
    \end{tabular}
      \caption[]{\label{fig::kss14nnlo}\sloppy $K$-factors for the
	gluon-fusion process.  Dashed: Standard Model --- Solid: \mssm{}
	(no stop mixing).  The narrow (red) band shows the uncertainty
	due to the missing \nnlo{} contribution in the effective vertex,
	the wide (green) band is the scale uncertainty (from
	\reference{Harlander:2003kf}).}
  \end{center}
\end{figure}

The dominant corrections to the Higgs production cross section originate
from real gluon emission~\cite{gghsoft}. Thus, it is possible to derive
a rather precise estimate of the \nnlo{} terms based on the \nnlo{}
result in the \sm{}~\cite{gghnnlo} and the \nlo{} effective Higgs-gluon
coupling~\cite{Harlander:2003bb}. In this way, the reduced scale
uncertainty of the \nnlo{} in the \sm{} directly carries forward to the
supersymmetric case. The result is shown in \fig{fig::kss14nnlo},
details can be found in~\reference{Harlander:2003kf}.

\section{Associated production with bottom quarks}
\label{sec::bbh}
\begin{figure}
  \begin{center}
    \leavevmode
    \begin{tabular}{c}
      \epsfxsize=\figsize
      \epsffile[110 265 465 560]{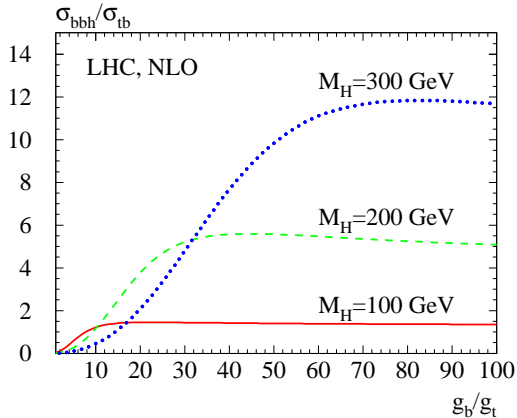}
    \end{tabular}
      \caption[]{\label{fig::tanbdep_lhc}\sloppy Relative contribution
	of the process $b\bar b\to H$ to the total \nlo{} cross section,
	compared to gluon fusion, as a function of the bottom Yukawa
	coupling. Note that for $\sigma_{b\bar bh}$ we use the running
	bottom Yukawa coupling with $m_b(m_b)=4.3$\,GeV, while for
	$\sigma_{tb}$ we use the on-shell expression with $m_b^{\rm
	OS}=5$\,GeV.  }
  \end{center}
\end{figure}
A large value of $\tan\beta$ (i.e., $\tan\beta\gtrsim m_t/m_b\approx
35$) not only leads to a significant contribution of virtual bottom
quarks to the gluon-Higgs coupling; it also brings in a new Higgs
production mechanism, namely $pp\to b\bar bH$. The relative importance
of both processes is shown in \fig{fig::tanbdep_lhc} as a function of
the bottom Yukawa coupling.

\begin{figure}
  \begin{center}
    \leavevmode
    \begin{tabular}{cc}
      \epsfxsize=8em
      \raisebox{.6em}{\epsffile[120 500 380 680]{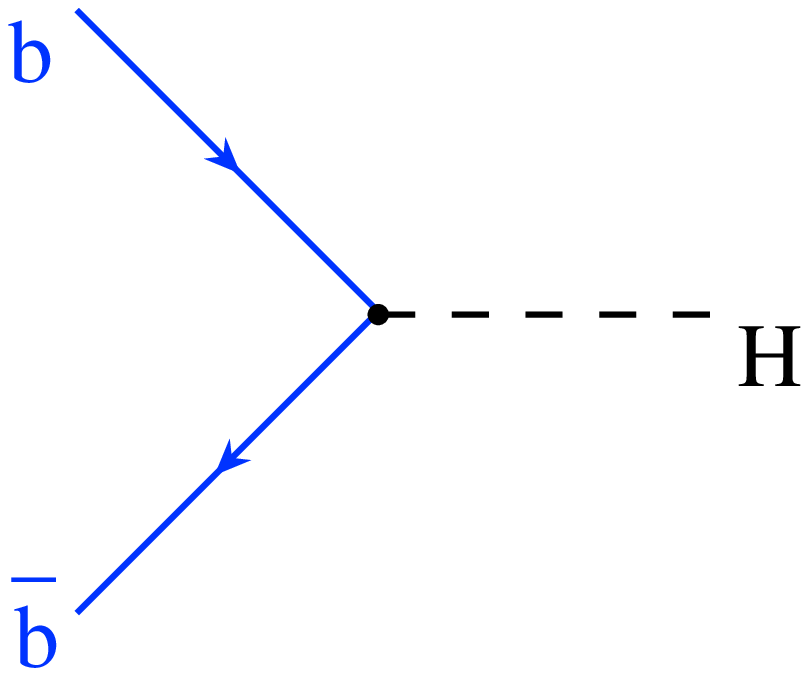}}
      &
      \epsfxsize=8em
      \raisebox{0em}{\epsffile[120 500 380 680]{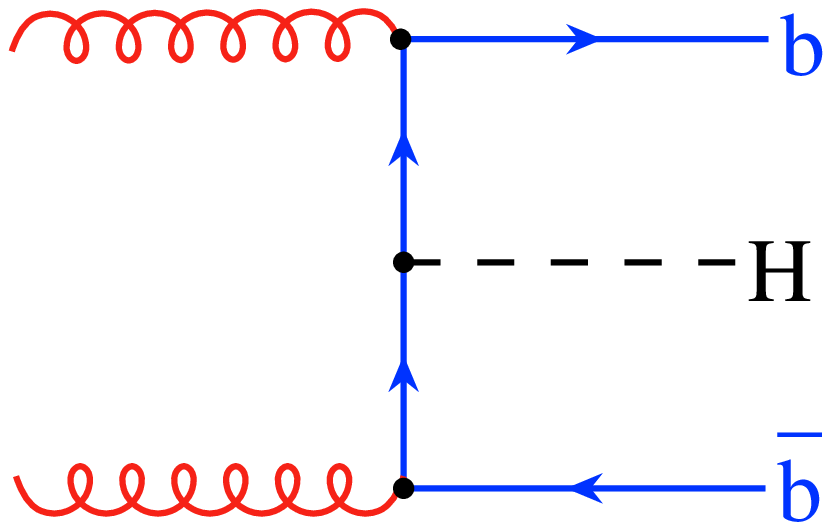}} \\
      $(a)$ & $(b)$
    \end{tabular}
      \caption[]{\label{fig::bbh}\sloppy Sample Feynman diagrams
        contributing to bottom annihilation $(a)$~at leading order ---
        $(b)$~at \nnlo{}.}
  \end{center}
\end{figure}

The \lo{} partonic Feynman diagram would be $gg\to b\bar bH$,
\fig{fig::bbh}\,$(b)$. However, integration over small transverse
momenta $p_{T,b}$ of the bottom jets leads to collinear logarithms $\sim
\ln(m_b/M_H)$.  Resummation of these logarithms can be achieved by
introducing bottom quark densities for the initial state hadrons, and
using $b\bar b\to H$ as the \lo{} partonic process (the two final state
bottom jets remain unobserved). The \lo{} contribution in this approach
neglects contributions from {\sl large} $p_{T,b}$; but they are
re-introduced through higher order {\abbrev QCD} corrections. In fact, a
\nnlo{} calculation~\cite{Harlander:2003ai} consistently combines the
all-order resummation of the low-$p_{T,b}$ region with the \lo{}
contributions from large $p_{T,b}$.  This is plausible because the
diagrams for $gg\to b\bar bH$ are naturally part of the \nnlo{}
contribution to the process $b\bar b\to H$ (see \fig{fig::bbh}).

The \nnlo{} result depends only very weakly on the renormalization and
factorization scales, thus providing a very precise prediction of the
inclusive rate. In addition, it supports the analyses of
\reference{bbhnlo} which suggest that the ``central'' choice of the
factorization scale for this process should be $\mu_F=M_H/4$.

Recently, the \nlo{} corrections for the {\sl exclusive} process became
available~\cite{Dittmaier:2003ej}.
In this case, the \lo{}
partonic process is indeed $gg\to b\bar bH$. When integrated over {\sl
all} bottom quark transverse momenta, the cross section is in good
agreement with the result from the bottom density
approach~\cite{Harlander:2003ai}, be it with much larger error bars.

\paragraph{Conclusions.}
The recent \nnlo{} results for Higgs production at hadron colliders have
brought confidence into the theoretical
predictions\footnote{Unfortunately, we could not include recent \nnlo{}
results on pseudo-scalar Higgs production~\cite{pseudo} and
Higgs-Strahlung~\cite{Brein:2003wg}.}.  The reduced uncertainties now
have to be achieved also in extended models. First steps have been done,
but there are still many problems to be solved that ask for technical
progress and new ideas.

\paragraph{Acknowledgments.}
It is a pleasure to thank Bill~Kilgore and Matthias~Steinhauser for
collaboration on many of the subjects discussed in this paper.
Furthermore, I would like to thank Michael~Spira for clarifying discussions.

\def\fortp#1#2#3{{\it Fortschr.~Phys.~}\jref{\bf#1}{#2}{#3}}
\def\jhep#1#2#3{{\small\it JHEP~}\jref{\bf #1}{#2}{#3}}
\def\npb#1#2#3{{\it Nucl.~Phys.~}\jref{\bf B #1}{#2}{#3}}
\def\plb#1#2#3{{\it Phys.~Lett.~}\jref{\bf B #1}{#2}{#3}}
\def\prd#1#2#3{{\it Phys.~Rev.~}\jref{\bf D #1}{#2}{#3}}
\def\prl#1#2#3{{\it Phys.~Rev.~Lett.~}\jref{\bf #1}{#2}{#3}}
\newcommand{\jref}[3]{{\bf #1}, #3 (#2)}
\newcommand{\arxiv}[1]{{\tt #1}}


\begin{thebibliography}{99}
%
% epj2003_ref.tex -- generated by sortref-2.3.3 on
% Fri Oct 31 19:19:21 CET 2003
%
%1
\bibitem{nlo}
S.~Dawson, \npb{359}{1991}{283}
%%CITATION = NUPHA,B359,283;%%
A.~Djouadi, M.~Spira, P.~Zerwas,
\plb{264}{1991}{440}.
%%CITATION = PHLTA,B264,440;%%

%\cite{Harlander:2003ai}
%2
\bibitem{Harlander:2000mg}
R.~Harlander,
\plb{492}{2000}{74}.
%%CITATION = HEP-PH 0007289;%%

%3
\bibitem{gghnnlo}
R.~Harlander, W.~Kilgore, \prl{88}{2002}{201801};
%%CITATION = HEP-PH 0201206;%%
C.~Anastasiou, K.~Melnikov,
\npb{646}{2002}{220};
%%CITATION = HEP-PH 0207004;%%
V.~Ravindran, J.~Smith, W.~van Neerven,
\npb{665}{2003}{325}.
%%CITATION = HEP-PH 0302135;%%

%4
\bibitem{gghresum}
S.~Catani, D.~de Florian, M.~Grazzini, P.~Nason,
\jhep{0307}{2003}{028};
%%CITATION = HEP-PH 0306211;%%
A.~Kulesza, G.~Sterman, W.~Vogelsang,
\arxiv{hep-ph/0309264}.

%5
\bibitem{Spira:1995rr}
M.~Spira, A.~Djouadi, D.~Graudenz, P.~Zerwas,
\npb{453}{1995}{17}.
%%CITATION = HEP-PH 9504378;%%

%6
\bibitem{higlu}
M.~Spira, \arxiv{hep-ph/9510347}.

%\cite{Harlander:2003bb}
%7
\bibitem{bill}
B.~Kilgore, talk at DPF\,2003.

%\cite{Harlander:2000mg}
%8
\bibitem{Harlander:2003bb}
R.~Harlander, M.~Steinhauser,
\plb{574}{2003}{258}.
%%CITATION = HEP-PH 0307346;%%

%\cite{Spira:1995rr}
%9
\bibitem{Dawson:1996xz}
S.~Dawson, A.~Djouadi, M.~Spira,
\prl{77}{1996}{16}.
%%CITATION = HEP-PH 9603423;%%


%10
\bibitem{gghsoft}
R.~Harlander, W.~Kilgore,
\prd{64}{2001}{013015};
%%CITATION = HEP-PH 0102241;%%
S.~Catani, D.~de Florian, M.~Grazzini,
\jhep{0105}{2001}{025}.
%%CITATION = HEP-PH 0102227;%%


%11
\bibitem{Harlander:2003kf}
R.~Harlander, M.~Steinhauser,
\arxiv{hep-ph/0308210}, {\it Phys.~Rev.}~{\bf D}, in print.
%%CITATION = HEP-PH 0308210;%%

%12
\bibitem{Harlander:2003ai} 
R.~Harlander, W.~Kilgore,
\prd{68}{2003}{013001}.
%%CITATION = HEP-PH 0304035;%%

%\cite{Boos:2003yi}
%13
\bibitem{bbhnlo}
F.~Maltoni, Z.~Sullivan, S.~Willenbrock,
\prd{67}{2003}{093005};
%%CITATION = HEP-PH 0301033;%%
E.~Boos, T.~Plehn,
\arxiv{hep-ph/0304034}.
%%CITATION = HEP-PH 0301033;%%

%\cite{Dittmaier:2003ej}
%14
\bibitem{Dittmaier:2003ej}
S.~Dittmaier, M.~Kr\"amer, M.~Spira,
\arxiv{hep-ph/0309204}.
%%CITATION = HEP-PH 0309204;%%
%15
\bibitem{pseudo}
R.~Harlander, W.~Kilgore,
\jhep{0210}{2002}{017};
%%CITATION = HEP-PH 0208096;%%
C.~Anastasiou, K.~Melnikov,
\prd{67}{2003}{037501}.
%%CITATION = HEP-PH 0208115;%%

%16
\bibitem{Brein:2003wg}
O.~Brein, A.~Djouadi, R.~Harlander,
\arxiv{hep-ph/0307206}
%%CITATION = HEP-PH 0307206;%%


\end{thebibliography}
\end{document}